\documentclass[11pt,twoside]{article}

\usepackage{asp2006}
\usepackage{epsf}

\def\gtrsim{\mathrel{\hbox{\rlap{\hbox{\lower4pt\hbox{$\sim$}}}\hbox{$>$}}}}
\def\lesssim{\mathrel{\hbox{\rlap{\hbox{\lower4pt\hbox{$\sim$}}}\hbox{$<$}}}}

\begin{document}

\title{Exoplanet-Induced Chromospheric Activity:
Realistic Light Curves from Solar-type Magnetic Fields}
\author{Steven R. Cranmer and Steven H. Saar}
\affil{Harvard-Smithsonian Center for Astrophysics,
60 Garden Street, Cambridge, MA 02138, USA}

\begin{abstract}
There is growing observational evidence for some kind of
interaction between stars and close-in extrasolar giant planets.
Shkolnik et al.\  reported variability in the chromospheric Ca H and
K lines of HD 179949 and $\upsilon$~And that seemed to be phased
with the planet's orbital period, instead of the stellar rotational
period. However, the observations also indicate that the
chromospheric light curves do not repeat exactly, which may be
expected for a planet plowing through a variable stellar
magnetic field. Using the complex solar magnetic field (modeled
with the Potential Field Source Surface technique) as a guide, we
simulate the shapes of light curves that would arise from planet-star
interactions that are channeled along magnetic field lines. We also
study the orbit-to-orbit variability of these light curves and how
they vary from solar minimum (i.e., a more or less axisymmetric
stretched dipole) to solar maximum (a superposition of many higher
multipole moments) fields. Considering more complex magnetic fields
introduces new difficulties in the interpretation of observations,
but it may also lead to valuable new diagnostics of exoplanet
magnetospheres.
\end{abstract}

\section{Introduction}

Many of the $\sim$200 known extrasolar planets are
``Hot Jupiters'' (i.e., giant planets with star-planet
distances less than about 0.1 AU) that have been identified
from perturbations in their host star's radial velocities.
Other ways in which a close-in extrasolar giant planet
(CEGP) can influence its star include tidal distortions and
magnetic interactions.
These effects can produce enhanced stellar flare activity
(Rubenstein \& Schaefer 2000),
chromospheric and coronal emission (Cuntz et al.\  2000), and
magnetospheric radio emission (Zarka et al.\  2001).
Tidal and magnetic interactions have been proposed to explain
similar phenomena observed in RS~CVn close binary systems (e.g.,
Simon et al.\  1980; Ferreira \& Mendoza-Brice\~{n}o 2005),
stars with possibly unseen companions exhibiting
``superflares'' (Schaefer et al.\  2000),
and even in the aurora of Jupiter where there are magnetic
connections to the inner Galilean moons (Clarke et al.\  2002).
We refer to Saar et al.\  (2004) for a recent review of
observations (or the lack thereof) of various kinds of
CEGP-star interactions and possible theoretical
explanations for these effects.

In this presentation we focus on the chromospheric
(Ca H \& K line) enhancements on a star due to magnetic
interactions between the stellar field and the CEGP's
magnetosphere.
We used the empirically derived solar magnetic field, as a function
of the 11-year solar cycle, as a proxy for the field geometry and
field strength of CEGP host stars.
We simulated light curves that represent the sum of the
rotational modulation of the star's own H\&K line emission and
the added emission from the small set of flux tubes connected
to the planet's magnetosphere.

\section{Observed Ca H \& K Enhancements}

Several early attempts to identify planet-induced
chromospheric emission (reviewed by Saar et al.\  2004)
were unsuccessful mainly because the observations were not
designed specifically for this purpose.
Shkolnik et al.\  (2003, 2005) used high-resolution
($\lambda / \Delta\lambda \sim 10^5$) Ca~II H \& K
spectroscopy to identify two cases of chromospheric enhancement that
are phased with the planet's orbit.
HD 179949 (F8 V, $P_{\rm rot} \approx 9$ days) has a CEGP
with $P_{\rm orb} = 3.092$ days, $M \sin i = 0.98 \,
M_{\rm Jup}$, and semimajor axis $d = 0.045$ AU.
The Ca K enhancement for
HD 179949 occurs at a phase shift $\Delta \phi$ of almost 0.2
{\em ahead} of the sub-planet point reaching stellar disk-center.
This raised suspicions that the star's magnetic field may be
swept back into a ``Parker spiral'' at the orbit of the CEGP,
but for a solar-type magnetic field and wind this is unlikely.
On the other hand, HD 179949 is more active than the Sun
(e.g., $F_{\rm X}/F_{\odot} \sim 20$) and of earlier spectral
type ($T_{\rm eff} \approx 6170$ K; Valenti \& Fischer 2005)
suggesting both the stellar magnetic structure and wind could
be quite different from the solar case.
Shkolnik et al.'s data from 2003 do not show the strong orbital
modulation of the 2001--2002 data, but there is still a
weak maximum at $\phi \sim 0.8$ in agreement with the earlier
epochs.
Similar data for $\upsilon$~And (F7 V, 
$P_{\rm rot} \approx 14$ days, 
$P_{\rm orb} = 4.6$ days) reveals a slightly weaker
H \& K amplitude, but a maximum with phase shift
$\Delta \phi \approx 0.5$, i.e., when the planet is behind the star
(see also Harrington et al.\  2006).

\section{Theoretical Explanations?}

The periodicity and strength of the HD 179949 and $\upsilon$~And
variations indicate that {\em magnetic interaction}
is the most likely mechanism.
Saar et al.\  (2004) found that other mechanisms such as tidal
distortions or Io-like inductance would have produced even stronger
modulations in other star-planet systems for which no planet-phased
variations have been found (e.g., $\tau$~Boo).
In addition, tidal distortions would have caused two peaks in the
light curve per orbit, rather than just one as is observed.
The energy released in the Ca H\&K enhancement ($\sim$10$^{27}$ erg/s)
is similar to that released in typical solar flares.
Ip et al.\  (2004) modeled CEGP magnetospheres and found that
this level of power generation can be expected from magnetic
reconnection in the most likely geometries.

McIvor et al.\  (2006) predicted the shapes of Ca H\&K light curves
consistent with magnetic field lines connecting a planet and an
inclined, stretched dipolar stellar field.
Phase shifts similar to that of HD 179949 can be reproduced by
misalignment between magnetic and rotation axes.
Also, Preusse et al.\  (2006) were able to model similar phase
shifts by assuming the CEGP to be inside the star's Alfv\'{e}n
radius (i.e., where the Alfv\'{e}n speed exceeds the outflow
speed; $r \lesssim 10 \, R_{\odot}$ for the solar wind) and that
the perturbations travel from the planet to the star along the
{\em longitudinally varying} inward Alfv\'{e}n characteristics.

Magnetic interaction models depend on the product of the relative
velocity between the two fields (known from the orbital dynamics),
the stellar magnetic field strength (which can be deduced from
Zeeman splitting or estimated from known activity scalings), and
the planetary magnetic field strength, which is not yet observed,
though modeling progresses (e.g.,
S\'anchez-Lavega 2004; Stevens 2005).

\section{Exploratory Models}

Our goal has been to produce more realistic simulations of
Ca~K light curves for a rotating solar-type star undergoing a CEGP
magnetic interaction.
We use modeled three-dimensional solar magnetic field configurations
(as a function of the 11-year solar cycle) to simulate both the
stellar and planetary light curve components.
Observed maps of the photospheric magnetic flux were extrapolated
into the extended corona using the Potential Field Source Surface
(PFSS) method.
The corona is assumed to remain current-free out to a spherical
``source surface''
at $R_{\rm ss} = 2.5 \, R_{\odot}$ where the nonradial field
components are set to zero, simulating the magnetohydrodynamic
(MHD) expansion of the solar wind (e.g.,
Schatten et al.\  1969; Altschuler \& Newkirk 1969;
Hoeksema \& Scherrer 1986; Wang \& Sheeley 1990).

The highest multipole components fall off the most rapidly with
increasing height, resulting in the photospheric field being much
more complex than the field at $r = R_{\rm ss}$.
Data and reconstruction coefficients were obtained for 11 solar
rotations (one per year over a solar cycle)
from Wilcox Solar Observatory.\footnote{%
http://sun.stanford.edu/$\sim$wso/}
Somewhat arbitrarily, we used the solar magnetic field from the
month of August in each of the years of solar cycle 22 (1986--1996);
i.e., the specific Carrington Rotations 1778, 1792, 1805, 1819,
1832, 1845, 1859, 1872, 1886, 1899, and 1913.

The following subsections describe how the stellar ({\S}~4.1)
and planetary ({\S}~4.2) light curve components were computed.

\subsection{Stellar Light Curve}

\begin{figure}[!ht]
\plotone{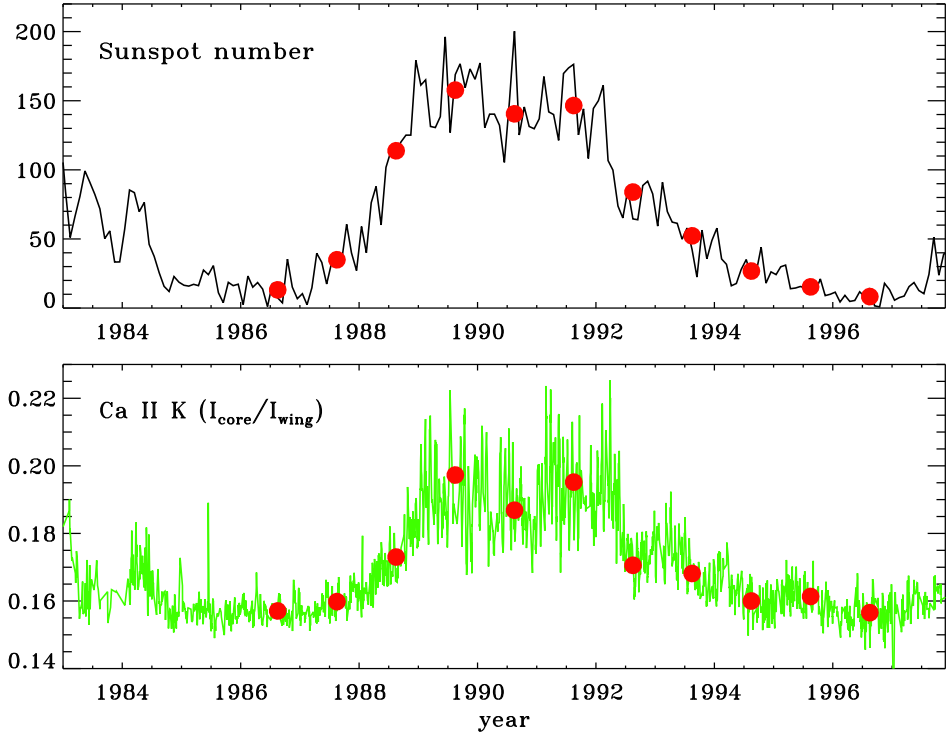}
\caption{{\em{Top:}} monthly raw sunspot number over cycle 22.
{\em{Bottom:}} Ca~II K-line ratio of core to wing intensity
(see text).  Red points in both panels define the 11 specific
solar rotations that we chose to model in detail.}
\end{figure}

Figure 1 shows the variation of sunspot number\footnote{%
http://sidc.oma.be/sunspot-data/}
and Ca K-line emission over solar cycle 22.
The larger number of sunspots at solar maximum corresponds
directly to the larger number of chromospheric plage and active
regions which enhance the disk-integrated K-line emission.
The Ca K-line data came from daily observations at the 
Sacramento Peak Observatory of the U.S. Air Force Phillips
Laboratory.\footnote{%
http://www.ngdc.noaa.gov/stp/SOLAR/ftpcalcium.html}
The Sac Peak $K_3$ index was divided by a fiducial
wing-to-continuum ratio of 0.40 in order to obtain the
same kind of core-to-wing intensity ratio used by
Shkolnik et al.\  (2003, 2005).
These values were then calibrated against the
hemisphere-averaged magnetic flux densities
$\langle B \rangle$ (measured at the solar surface) over
the modeled epochs.
We performed a similar fit between these two quantities as
did Schrijver et al.\  (1989), and found
\begin{equation}
  (I_{\rm core} / I_{\rm wing} ) \, = \,
  0.13 \, + 0.024 \langle B \rangle^{0.63} 
\end{equation}
where $\langle B \rangle$, measured in Gauss, was computed
from the PFSS models over the 11 modeled rotations at times
corresponding to the K-line observations.
This relation was then used to compute the theoretical
stellar-rotation light curves in Figures 3 and 4
from the ``exactly known'' photospheric magnetic fields.

\subsection{Planetary Light Curve}

To model the enhancement due to planet-star coupling, we assumed
the power released in magnetic interaction scales as
\begin{equation}
  P \, \sim \, \frac{B_{\ast} B_{\rm P}}{8\pi}
  \, ( \pi r_{\rm mag}^{2} )
  \, V_{\rm rel}
\end{equation}
where the stellar magnetic field ($B_{\ast}$) and planetary field
($B_{\rm P}$) are measured at the planet's magnetosphere, which has
a projected area $\pi r_{\rm mag}^{2}$ intercepted
by the stellar field (e.g., Saar et al.\  2004).
Later, we use $r_{\rm mag} \approx 5 \, R_{\rm Jup}$, which was
also assumed by Ip et al.\  (2004).
The two magnetic fields move through one another at
relative velocity $V_{\rm rel}$ ($\sim$100 km/s for HD 179949).
We assumed that $B_{\rm P}$, $r_{\rm mag}$, and $V_{\rm rel}$ remain
fixed over multiple rotations and the activity cycle, and thus
that {\em time variations} in $P$ are sensitive only to $B_{\ast}$.

\begin{figure}[!ht]
\plotone{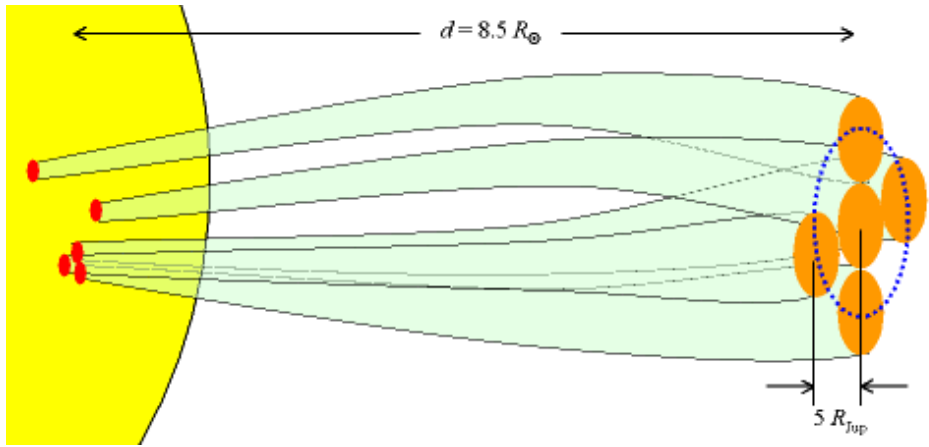}
\caption{Cartoon illustrating the coarse array of magnetic flux
tubes traced down from the CEGP magnetosphere to the stellar
surface.  Not to scale.}
\end{figure}

For each of the 11 epochs, we created a fine grid of planetary
longitudes (300 points around the circular orbit) defined relative
to the fixed Carrington longitude coordinates of the modeled
magnetic field.
For each planetary position, we traced an envelope of 5
flux tubes from the planet's orbit ($d > R_{\rm ss}$) down to
the stellar surface.
Figure 2 shows the cross-like configuration of flux tubes that
we used to resolve the magnetosphere; note that the pattern of
flux tube ``footpoints'' on the stellar surface is likely to
have a more complicated shape.
We used more than one flux tube so that ``bifurcation'' events
(when the finite-sized magnetosphere connects down to two widely
separated areas on the stellar surface) could be at least
approximately resolved.
Future work should of course use a more finely sampled envelope
of flux tubes so that the detailed shapes of the resulting
star spots can be computed.

The mapped flux tubes were then input into a time-resolved model
of both the planet's orbit and the star's rotation.
The instantaneous angular positions of the star, planet, and
observer were tracked, and the angle $\theta$ (measured between
the stellar surface normal at the flux-tube footpoint and the
observer's line of sight) was computed as a function of time.
The relative contribution from each flux tube was thus
assumed to be proportional to the stellar field strength
at the magnetosphere ($B_{\ast}$) modulated by the spot's
projection factor; i.e.,
\begin{equation}
  (I_{\rm core} / I_{\rm wing}) \, \propto
  \, \sum \, \left( B_{\ast} \, \cos\theta \right)
  \,\,\,\,\,
  \mbox{(summed over 5 footpoints)}
\end{equation}
The above procedure contains the assumption that the magnetic
``communication'' between star and planet is essentially
instantaneous.
This would be a valid assumption if the energy were carried
by, e.g., suprathermal particles as in strong solar flares,
but not if the energy had to propagate along more slowly as
an MHD wave (see Preusse et al.\  2006).

We adopted dynamical parameters roughly analogous to HD 179949
(e.g., $P_{\rm rot} = 2.9 P_{\rm orb}$, not an exact multiple),
and we set the overall
normalization for $I_{\rm core} / I_{\rm wing}$ to resemble
the observed amplitude (Shkolnik et al.\  2005).
This arbitrary normalization is of course a weak point of
the present models, but without a specific predictive model of
the magnetic interaction not much more was possible.

\section{Results}

\begin{figure}
\plotone{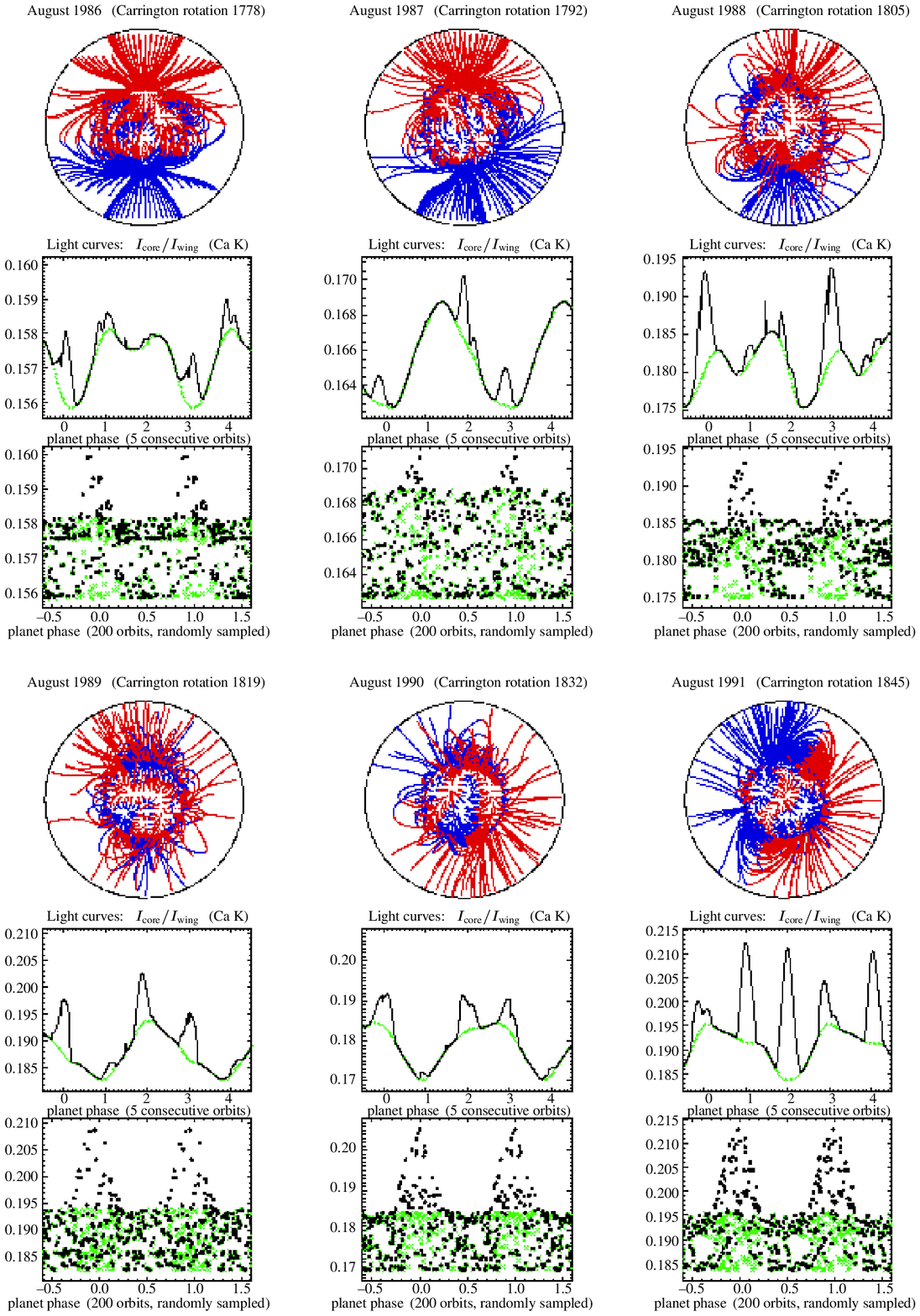}
\caption{For years 1986--1991, the top panel shows
the star's magnetic field out to $R_{\rm ss}$, with positive
[negative] polarity in blue [red].
The middle panel shows a representative light curve over 5
planetary orbits, and the bottom panel shows a coarser random
sampling over 200 orbits all phased together.
In both, the full star$+$planet light curve is in
black (solid curves, filled circles), and the stellar
rotational component only is
in green (dotted curves, X's).}
\end{figure}

\begin{figure}
\plotone{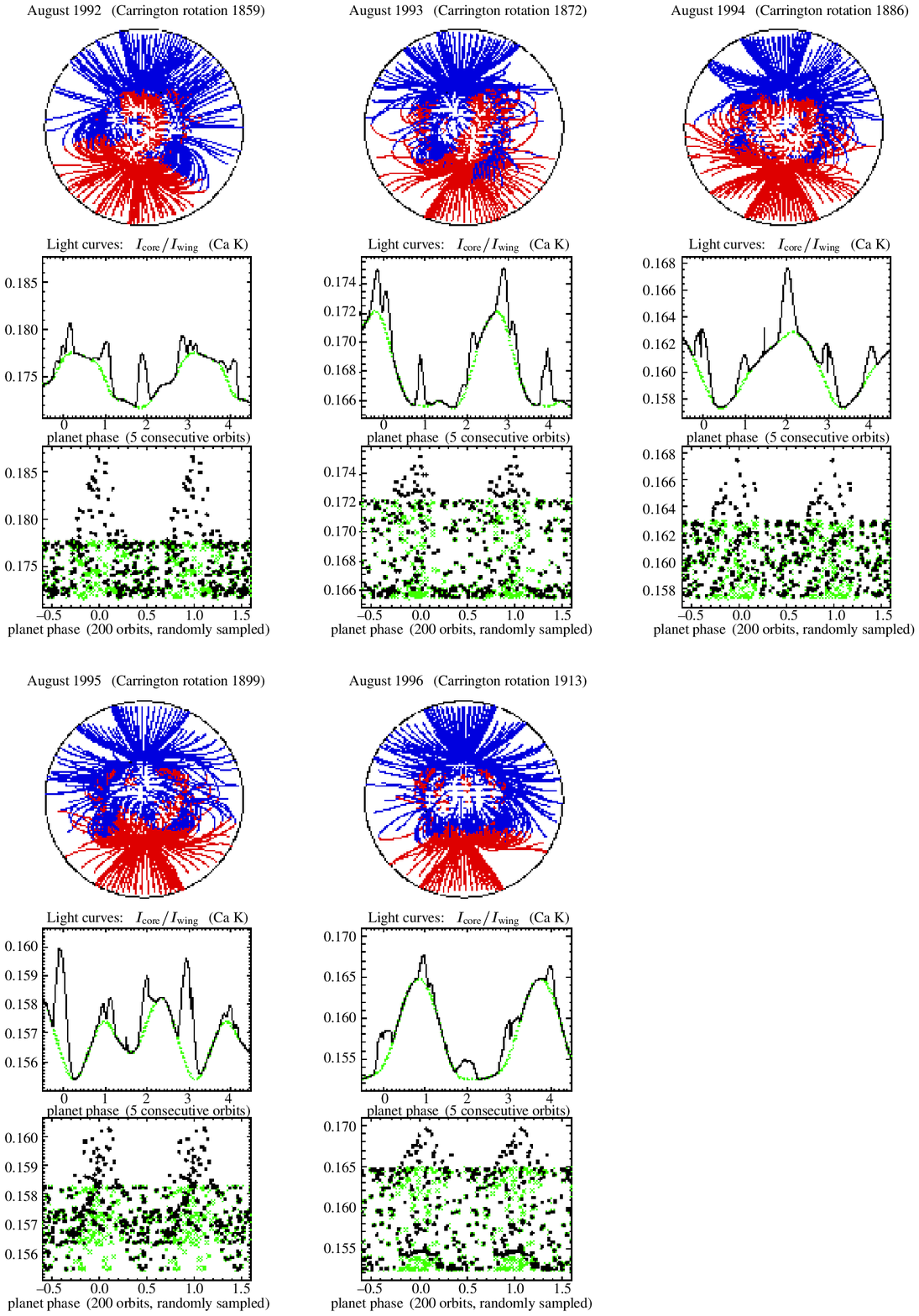}
\caption{For years 1992--1996, the top panel shows
the star's magnetic field out to $R_{\rm ss}$, with positive
[negative] polarity in blue [red].
The middle panel shows a representative light curve over 5
planetary orbits, and the bottom panel shows a coarser random
sampling over 200 orbits all phased together.
In both, the full star$+$planet light curve is in
black (solid curves, filled circles), and the stellar
rotational component only is
in green (dotted curves, X's).}
\end{figure}

Figures 3 and 4 show the magnetic field configurations and
light curves for the 11 modeled epochs.
The overall polarity ``flip'' is evident over the 11-year
solar cycle.
Both the complete light curve (star plus planet) and just the
stellar rotational component are shown in order to clearly
see the planetary enhancement.
The light curve is shown in detail over five planetary orbits.
The planetary enhancements generally occur when the planet is
in front of the star ($\phi \sim 0$) but occasional planetary
enhancements occur at other phases, or are absent at phase zero.
Rapid changes sometimes occur over less than 0.05 in phase
(i.e., over just several hours in time).
We also show a much coarser kind of ``light curve'' that was
randomly sampled with 200 points spread over a time-span of
200 planetary orbits.
In these plots, the stellar rotation light curve is not resolved
except as a fuzzy ``band'' which the planetary enhancement tends
to exceed around phase zero.

\begin{figure}[!ht]
\plotone{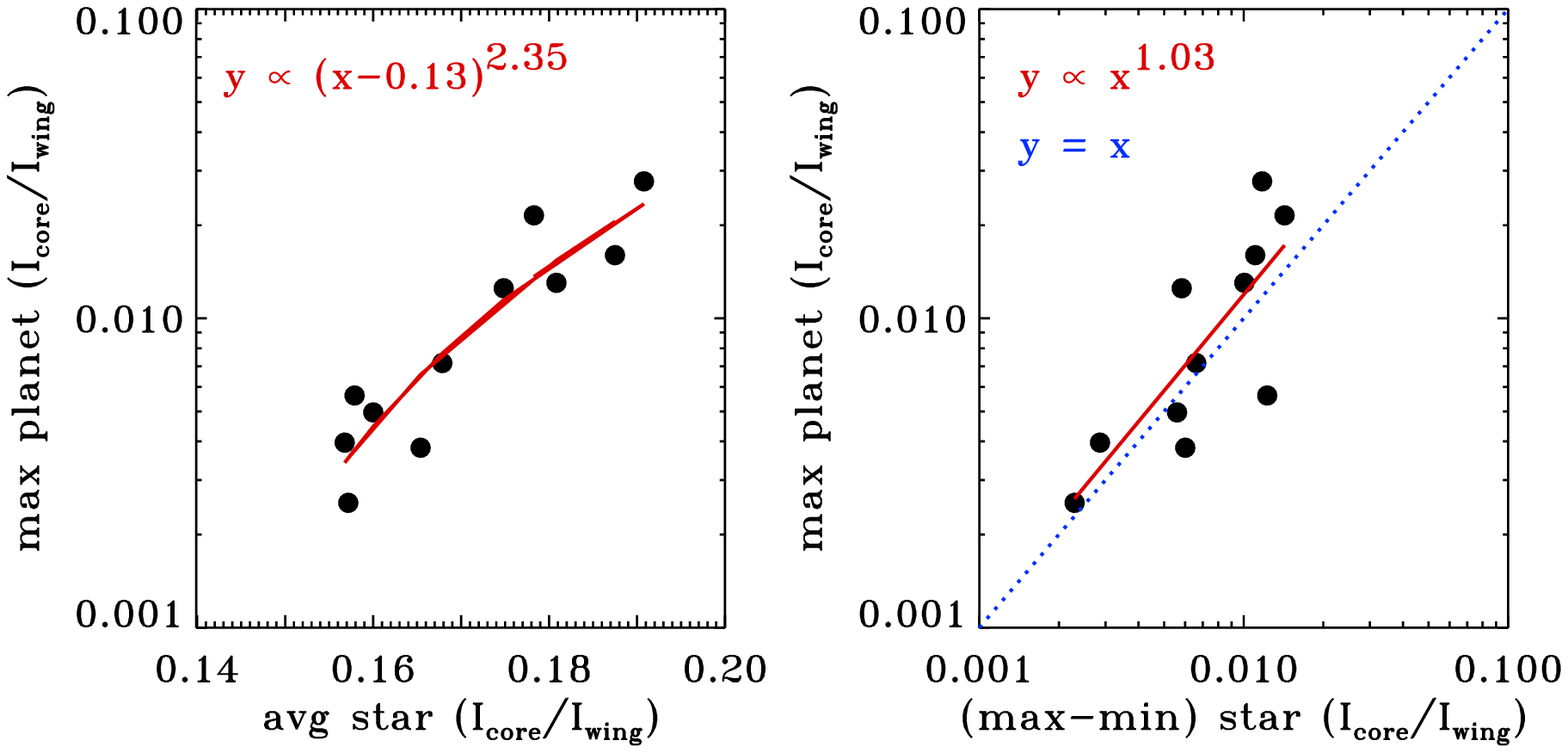}
\caption{Correlations between statistical properties of the
stellar (abscissa) and planetary (ordinate) light curves over
the 11 modeled epochs.}
\end{figure}

Because the amplitudes of {\em both} the stellar-rotation light
curve and the planetary enhancements are larger at solar maximum
than at solar minimum, there is no systematic trend for the
planetary enhancement to ``poke its head above the fuzz''
more at either maximum or minimum.
Thus, there does not seem to be any preferred cycle-phase
for the planetary enhancement to be more or less distinguishable
from the stellar rotational modulation.

Figure 5 shows some statistical properties of the light curves,
with one point in each panel characterizing the variations over
each of the 11 epochs.
The minimum, maximum, and average intensity ratios
($I_{\rm core} / I_{\rm wing}$) were determined for the separate
stellar and planetary light curves and correlated with one another.
The tendency for the amplitudes of the stellar and planetary
light curves to rise and fall together was discussed above
and is evident in the direct correlations.
Note specifically the nearly linear relationship between the
maximum planet enhancement and the total (max $-$ min) amplitude of
the stellar light curve.

\section{Conclusions}

Some of the more interesting features of the modeled light
curves are summarized here:
\begin{enumerate}
\item
Because of the complex nature of the multipole fields,
the modeled light curves do not repeat exactly
from orbit to orbit, and sometimes the planetary enhancement
seems to disappear altogether.
This may be a possible explanation for the 2003 disappearance of
the strong orbital modulation seen in 2001--2002 for HD 179949
(Shkolnik et al.\  2005).
\item
The planetary enhancements are often
non-monotonic in shape (i.e., multiply peaked).
This occurs because the ``spots'' on the stellar surface that
connect to the planet's magnetosphere may be bifurcated or
substantially swept around in longitude and/or latitude.
\item
For sparsely sampled data, the inferred phase shift
between light-curve maximum and planetary meridian-passage
may range between $-0.2$ and $+0.2$.
\end{enumerate}
Better observational phase coverage (even using multiple
observatories to obtain round-the-clock time series)
over consecutive orbits and stellar rotations
would yield much better tests of the magnetic interaction
paradigm.
More work needs to be done, of course, to model how the stellar
magnetic fields differ from the fiducial solar case modeled here.
Additional kinds of observations, such as X-rays that may be
phased with the planet's orbit (Saar et al., these proceedings),
will help constrain the dynamics and energetics of the
interaction.

\acknowledgements This work was supported by the National
Aeronautics and Space Administration (NASA) under grants
{NNG\-04\-GE77G} and {NNG\-04\-GL54G}
to the Smithsonian Astrophysical Observatory.

\end{document}